\documentclass[twocolumn,showpacs,preprintnumbers,amsmath,amssymb,prb,aps]{revtex4-1}
\usepackage{epsfig}
\usepackage{epstopdf}
\usepackage{graphicx}
\usepackage{dcolumn}
\usepackage{bm}
\usepackage{textcomp}
\usepackage{array}
\usepackage{subcaption}
\usepackage{booktabs}
\usepackage{natbib}
\usepackage{url}
\setcitestyle{numbers,square}

\begin{document}

\title{First-principles electronic structure, phonon properties, lattice thermal conductivity and prediction of figure of merit of FeVSb half-Heusler }
\author{Shivprasad S. Shastri}
\altaffiliation{Electronic mail: shastri1992@gmail.com}
\author{Sudhir K. Pandey}
\altaffiliation{Electronic mail: sudhir@iitmandi.ac.in}
\affiliation{School of Engineering, Indian Institute of
Technology Mandi, Kamand - 175075, India}


\begin{abstract}
In this work, we have studied the electronic structure of a promising thermoelectric half-Heusler FeVSb using FP-LAPW method and SCAN meta-GGA including spin-orbit coupling. Using the obtained electronic structure and transport calculations we try to address the experimental Seebeck coefficient $S$ of FeVSb samples. The good agreement between the experimental and calculated $S$ suggests the band gap could be $\sim$0.7 eV. This is supported by the obtained mBJ band gap of $\sim$0.7 eV. Further, we study and report the phonon dispersion, density of states and thermodynamic properties. The effect of long range Coulomb interactions on phonon frequencies are also included by non-analytical term correction. Under quasi-harmonic approximation, the thermal expansion behaviour upto 1200 K is calculated. Using the first-principles anharmonic phonon calculations, the lattice thermal conductivity $\kappa_{ph}$ of FeVSb is obtained under single-mode relaxation time approximation considering the phonon-phonon interaction. At 300 K, the calculated $\kappa_{ph}$ is $\sim$18.6 W$m^{-1}K^{-1}$ which is higher compared to experimental value. But, above 500 K the calculated $\kappa_{ph}$ is in good agreement with experiment. A prediction of figure of merit $ZT$ and efficiency for p-type and n-type FeVSb is made by finding out optimal carrier concentration. At 1200 K, a maximum $ZT$ of $\sim$0.66 and $\sim$0.44 is expected for p-type and n-type FeVSb, respectively. For p-type and n-type materials, maximum efficiency of $\sim$12.2 \% and $\sim$6.0 \%  are estimated for hot and cold temperature of 1200 K and  300 K, respectively. A possibility of achieving n-type and p-type FeVSb by various elemental doping/vacancy is also discussed. Our study is expected to help in further exploring the thermoelectric material FeVSb.

\end{abstract}

\maketitle

\section{Introduction} 
Thermoelectric materials are the solids in which conversion between heat and electricity can be observed. This property of thermoelectric materials has drawn researchers to study and utilize them in power generation or cooling applications. These materials can be used to generate useful electricity from the waste heat released at many appliances, heat engines or industries. This helps in the recovery of waste heat in the form of useful electricity thereby reducing damage to the environment \cite{compintro,tritt}. Being solid state energy conversion devices, they are noise and vibration free, easily portable, scalable and less harmful to the environment compared to the fossil fuel based power generators. They have been finding applications where compared to cost and efficiency, energy availability and reliability are more important. The extensive application of thermoelectric materials in power generation is constrained by their relatively lower conversion efficiency \cite{dressel}. 

Improving efficiency depends on the materials figure of merit apart from the working temperature region of the material. A quantity which assigns a number according to the quality of a material for thermoelectric application is given by the dimensionless figure of merit $ZT$ as  \cite{snydercomplex,mgdintro},
\begin{equation}
ZT=\frac{S^{2}\sigma T}{\kappa}.
\end{equation}
Here, $S$ is the Seebeck coefficient, $\sigma$ is electrical conductivity, $T$ is absolute temperature and  $\kappa$ is total thermal conductivity which is sum of electronic ($\kappa_{e}$) and lattice ($\kappa_{ph}$) part of thermal conductivity. The figure of merit can be enhanced by improving the power factor ($S^{2}\sigma$) and reducing the thermal conductivity. But, obtaining higher $ZT$ is constrained since the transport coefficients are counter related and this leads to a task of optimization among these coefficients to get higher $ZT$ \cite{snydercomplex,mahanbest}. Therefore, to achieve that optimization one is lead to the problem of searching new materials, finding suitable doping elements to modify electronic structure and/or to reduce thermal conductivity. Till now, there are a few number of \textit{state-of-the-art} thermoelectric materials which work normally in lower and mid temperature range with high efficiency \cite{snydercomplex}. So, it is desirable to find materials which can be used in high temperature applications that show higher $ZT$. For such a study computational approach using first-principle methods is helpful with relatively less resource and time consuming.

In the search of thermoelectric materials for high temperature applications Heusler family of compounds are seeking attention in recent years. The presence of narrow bands with large Seebeck coefficient, moderate electrical conductivity, high melting point, mechanical strength and thermal stability are the features which make them desirable for thermoelectric applications \cite{felser,hhreview1}. Some of the half-Heuslers such as MNiSn and MCoSb (M = Ti, Zr, Hf) are found to be promising candidates from many experimental and theoretical studies \cite{hhreview1,hhreview2,hhreview3,hhreview4}. An understanding of electronic structure, band gap, phonon properties and lattice thermal conductivities of these compounds are given in number of earlier works \cite{hhreview1,hhreview3,rabe,katre,kappazrnisn,kappamcosb,kappatinisn,paper4}. Apart from these half-Heuslers, recently FeVSb based compounds are also gaining importance as promising candidates due to their high Seebeck coefficient and power factor \cite{hhreview1,hhreview2}.	

The FeVSb compound with 18 valence electrons per formula unit is a semiconducting half-Heusler. This nature makes it useful to explore for thermoelectric applications \cite{nanda2003}. The 18 valence electrons per formula unit also suggests that FeVSb is non-magnetic in nature from the Slate-Pauling rule. Nanda \textit{et al.} have performed a detailed study of the electronic structure, bonding and magnetism in a series of half-Heusler compounds including FeVSb \cite{nanda2003}. Their study using FP-LMTO-GGA method showed that the atomic arrangement leading to the semiconducting gap in the paramagnetic phase of FeVSb is the stable ground state structure. The obtained indirect band gap of FeVSb was found to be 0.36 eV.  From the TB-LMTO-ASA electronic structure analysis, the  gap at the Fermi energy was found to result from the covalent hybridization of the transition elements Fe and V in this work \cite{nanda2003}. The experimental thermoelectric properties of FeVSb was reported by Young \textit{et al.} with a room temperature Seebeck coefficient of $\sim$ -70 $\mu VK^{-1}$ \cite{young}. The study of experimental Seebeck coefficient and electronic structure of FeVSb and its doped compounds are carried out by Jodin \textit{et al.} \cite{jodin}. The room temperature Seebeck coefficient in this work for the FeVSb sample is $\sim$ -106 $\mu VK^{-1}$ with a maximum value ($\sim$ -150 $\mu VK^{-1}$) at about 600 K. The KKR-LDA calculations of electronic structure in this work showed an indirect gap of 0.46 eV \cite{jodin}. The substitution of Nb is found to be useful to decrease the thermal conductivity of FeVSb \cite{fu2012}. The reported value of Seebeck coefficient at room temperature by Fu \textit{et al.} are $\sim$ -215 $\mu VK^{-1}$ \cite{fu2012} and $\sim$ -200 $\mu VK^{-1}$ \cite{fu2013}, for undoped FeVSb. These values are higher compared to the experimental values of Ref. \cite{young} and \cite{jodin} mentioned before. They obtained a maximum  $ZT$  of $\sim$ 0.25 for the undoped FeVSb at 550 K depending upon the synthesis procedure \cite{fu2013}. 

Yamamoto \textit{et al.} performed an experimental and theoretical study of thermoelectric properties of FeVSb \cite{yamamoto}. The Seebeck coefficient of $\sim$ -165 $\mu VK^{-1}$ is obtained at room temperature. The obtained Seebeck coefficient in this work could be explained using the combined electronic structure from FP-LAPW GGA with an obtained band gap of 0.36 eV and Boltzmann transport calculations \cite{yamamoto}. The band gap is an important quantity in obtaining a good agreement between the experimental and calculated Seebeck coefficient. However, for FeVSb experimentally reported band gap from optical measurements is not found in literatures.  Also, the nature and magnitude of the Seebeck coefficient reported in the work of Fu \textit{et al.} \cite{fu2012,fu2013} are found to be different compared to that of the works in Ref. \cite{yamamoto}  and \cite{young}. So, In this work, we try to understand the experimental Seebeck coefficient of two samples \cite{fu2012,fu2013} using SCAN meta-GGA and band gap from mBJ. Studying the phonon properties of a thermoelectric materials is important to understand and enhance the thermal expansion or thermal conductivity. But, there is no reported theoretical study of phonon dispersion, thermal expansion or thermal conductivity of the promising thermoelectric material FeVSb in the literature to the best of our knowledge. This leads us to further study the band gap, various Seebeck coefficient, electronic structure and phonon properties of FeVSb.

Considering the above mentioned aspects, in this work we study the electronic structure of FeVSb using FP-LAPW method including spin-orbit coupling (SOC) with SCAN meta-GGA. The SCAN functional is found to be more appropriate to describe transport properties for Heusler compound relative to other functionals  \cite{shamim}. The mBJ functional \cite{mbj} is also used to get band gap which is $\sim$0.7 eV in case of FeVSb. Using the electronic dispersion with SOC and mBJ gap we try to explain the experimental Seebeck coefficients of two FeVSb samples. Then, the phonon dispersion, density of states and thermodynamical properties are calculated. The contribution to phonon modes from vibrations of different atoms are discussed. While calculating the phonon properties the effect of long range Coulomb interactions is considered by including nonanalytical term correction wherever significant. The thermal expansion behaviour is obtained under quasi-harmonic approximation by calculating linear thermal expansion coefficient with temperature. From \textit{ab-initio} anharmonic lattice dynamics calculations the lattice thermal conductivity of FeVSb is obtained by considering phonon-phonon interaction under single mode relaxation time approximation in 300 - 1200 K. The calculated lattice thermal conductivity at 300 K is $\sim$18.6 W$m^{-1}K^{-1}$ and it reaches value of $\sim$7.8 W$m^{-1}K^{-1}$  at 700 K which is close to experimental value of $\sim$7.0 W$m^{-1}K^{-1}$ (700 K), while the 300 K value is  $\sim$12.1 W$m^{-1}K^{-1}$. Further, optimal carrier concentration, corresponding figure of merit and efficiency are predicted considering three values of electronic relaxation time ($\tau$). For p-type and n-type FeVSb, 	maximum figure of merit of $\sim$0.66 and $\sim$0.44 at 1200 K are predicted. Correspondingly,  efficiency of $\sim$12.2 \% and $\sim$6.0 \% are obtained for cold and hot end temperature of 300 K and 1200 K, respectively.

\section{Computational details}
For the DFT calculations in this work WIEN2k \cite{wien2k} is used which is based on the full-potential augmented plane wave (FP-LAPW) method of DFT. The meta-GGA functional SCAN \cite{scan} is used for exchange and correlation (XC) part. The muffin-tin sphere radii ($R_{MT}$) of 2.17, 2.11 and 2.17 bohr are chosen for Fe, V and Sb atoms, respectively. In the total energy calculations a convergence criteria of 10$^{-4}$Ry/cell is used to reach self-consistency. A k-point mesh of 50 x 50 x 50 is used in order to get converged transport properties further. The electronic transport coefficients are calculated using BoltzTraP program \cite{boltztrap} based on Boltzmann transport theory. 

To calculate the phonon properties under harmonic approximation the forces on atoms are obtained using  WIEN2k. Using these forces, further second force constants and phonon properties are calculated under supercell and finite displacement method from PHONOPY \cite{phonopy}. In order to calculate forces on atoms a supercell (with 96 atoms) of the conventional unit cell of size 2 x 2 x 2 is constructed. In the calculation of forces XC functional SCAN is used. A k-point mesh of 5 x 5 x 5 is used to sample the Brillouin zone of the supercell. For the convergence of forces a criteria of 0.1 mRy/bohr is used. The thermal expansion is calculated under quasi-harmonic approximation \cite{phonopy} as implemented in PHONOPY. 

To calculate the lattice thermal conductivity, the forces on atoms are obtained using the project augmented wave (PAW) method of DFT as implemented in the ABINIT package \cite{abinit}. The local density approximation of the Perdew-Wang 92 type is used for the XC part \cite{lda92}. The PAW datasets are taken from the PseudoDojo \cite{pseudo}. A plane wave energy cutoff of 25 Ha is used and twice of this value is used for the PAW energy cutoff. A force tolerance criteria of 5x10$^{-8}$ Ha/bohr is used in scf calculations. A supercell of the the same size as used in the WIEN2k force calculations is also used here. The Brillouin zone of supercell is sampled by a 4 x 4 x 4 k-mesh.  

The second and third order force constants are obtained under supercell method with a finite displacement of 0.06 bohr in PHONO3PY \cite{phono3py}. The lattice thermal conductivity is calculated using PHONO3PY under single mode relaxation time approximation. A high q-point sampling mesh of 23 x 23 x 23 is used to get lattice thermal conductivity. In order to avoid the large number of supercell calculations considering the available computational resource, a real-space cutoff distance of 7.8 bohr is fixed ensuring three neighbor atoms interaction in this distance.

\section{Results and Discussion}
The half-Heusler under study, FeVSb has cubic $C1_b$ structure with space group $F\overline{4}3m$ (no. 216) \cite{felser,nanda2003}. In this structure, the Fe, V and Sb atoms occupy the Wyckoff positions $4c(1/4, 1/4, 1/4)$, $4a(0, 0 , 0)$  and $4b(1/2, 1/2, 1/2)$, respectively \cite{nanda2003}. Initially, the crystal structure is built using the experimental lattice constant of 5.820 \AA   \cite{fu2013}. The equilibrium ground state lattice constant is found out by fitting BM-EOS \cite{birch} to the energy vs. volume curve. The lattice constant thus obtained from SCAN functional is 5.733 \AA. Further electronic and phonon calculations are carried out using this lattice constant. 
\subsection{\label{sec:level2}Electronic structure}
The calculated electronic dispersion of FeVSb using SCAN functional is shown in Fig. 1 (a). The energy values in the figure are with respect to the Fermi level $E_{F}$. The Fermi level is set at the middle of the band gap in the plot. We also included the spin-orbit interaction to see the effect on the electronic energy levels close to the $E_{F}$ which mainly participate in the transport. The electronic dispersion so obtained with SOC is also shown in the figure along with the one from no-SOC.  As can be seen from the figure, FeVSb is an indirect band gap semiconductor with band gap of $\sim$0.33 eV (no-SOC). The electronic structure of FeVSb half-Heusler in its ground state is studied in some of the earlier works and it is reported to be an indirect band gap semiconductor. The obtained indirect band gap in the work of Nanda \textit{et al.} \cite{nanda2003}  from  FP-LMTO-GGA is 0.36 eV.  Yang \textit{et al} \cite{yang} and Jodin \textit{et al.} \cite{jodin} have reported band gaps of 0.32 eV and  0.46 eV, from PAW-GGA and KKR-LDA methods, respectively.  The calculated value of band gap from SCAN in our work is close the works of Ref. \cite{nanda2003,yamamoto} and \cite{yang}. We also calculated band gap using mBJ functional since it is known to give accurate values of band gaps for semiconductors \cite{mbj}. The obtained value of band gap in this case is $\sim$0.7 eV. This value is quite higher compared to the band gap obtained from other functionals/methods.

 In FeVSb, the valence band maxima (VBM) at $L$ and $W$-point are double degenerate and the conduction band minimum (CBM) at $X$-point is non-degenerate before including SOC.  The inclusion of the spin-orbit interaction has only slightly reduced the band gap to $\sim$0.32 eV compared to that of no-SOC. Also, the degeneracy of VBM at $L$ and $W$ points are lifted due to SOC. The effect of SOC and lifting of degeneracy around VBM ($L$ and $W$ points) are presented more clearly in Fig. 1 (b). In Fig. 1 (b) this lifting of bands from degeneracy at the VBM can be observed. The separation of bands at VBM indicates that the dynamics of holes in these bands which mainly contribute in the transport is going to be different compared to the case of no-SOC. Therefore, the effect of SOC is considered in our calculation as this would give more accurate and complete description of electronic structure and thus contribution to transport properties from holes. But, the effect of SOC around the conduction band bottom at $X$-point is found to be very less relative to that of VBM. Now, due to SOC there are four bands at VBM and two bands at the CBM to describe the transport in FeVSb. Normally, in a semiconductor, the transport  is mainly determined by the electrons and holes in the vicinity of the CBM and VBM. Effective mass of the carriers around these band extrema is an important quantity to explain the transport in a semiconductor. Therefore, the effective mass tensors for these bands at $L$-point (VBM) and $X$-point (CBM) are calculated to further study the temperature dependence of transport properties. The obtained eigenvalues \textit{m$_{1}$}, \textit{m$_{2}$}, \textit{m$_{3}$} of these effective mass tensors are tabulated in Table I. These values can be used to estimate the shift in the chemical potential with the temperature in a nondegenerate semiconductor. The calculated eigenvalues suggest that the effective mass of holes is higher compared to that of the electrons at the band extrema in FeVSb. 
\begin{figure*}
\includegraphics[width=16cm, height=8cm]{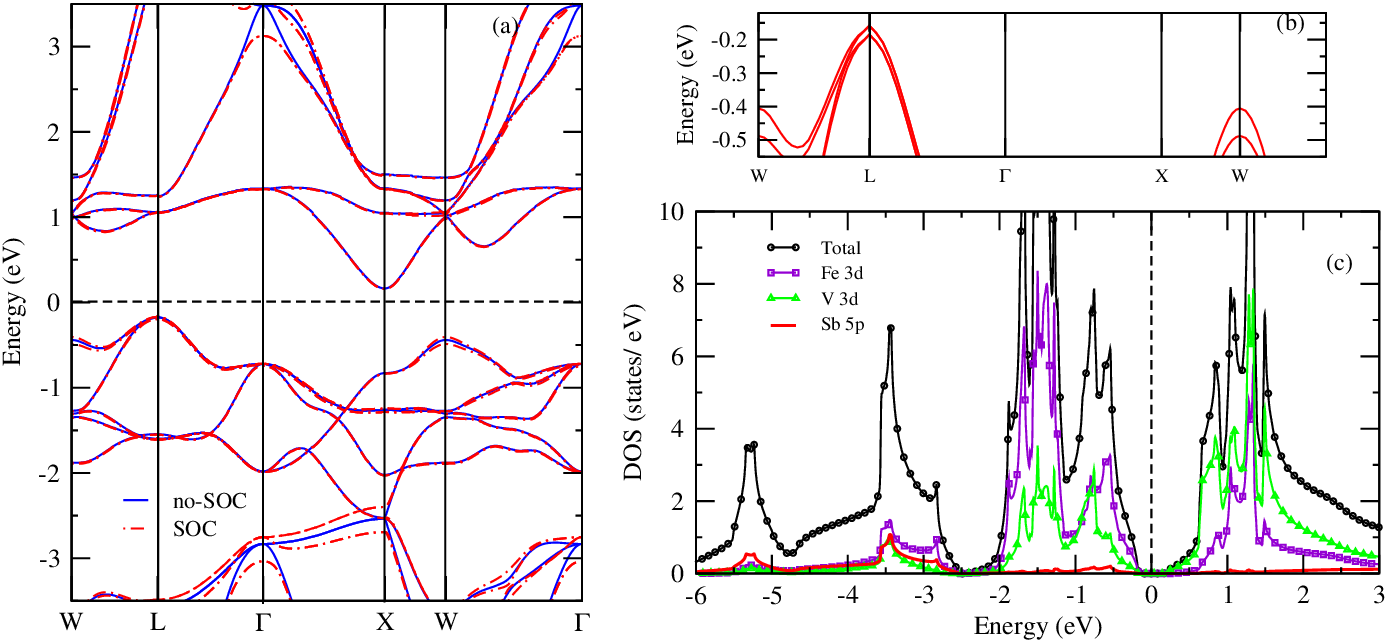} 
\caption{(a) Electronic dispersion of FeVSb without SOC (no-SOC) and including SOC. (b) The effect of SOC on the valence band maxima (VBM) at $L$ and $W$-point. (c) The electronic total DOS in states/eV/f.u. and partial DOS in states/eV/atom for FeVSb.}
\end{figure*}

\begin{table}[b]
\caption{\label{tab:table1}%
The eigenvalues m$_{1}$, m$_{2}$, m$_{3}$ of effective mass tensors of charge carriers at CBM ($X$-point) and VBM ($L$-point).
}
\begin{ruledtabular}
\begin{tabular}{lccc}
\textrm{}&
\textrm{\textit{m$_{1}$}}&
\textrm{\textit{m$_{2}$}}&
\textrm{\textit{m$_{3}$}}\\
\colrule
    &-0.714 &-0.711 &-0.621\\
VBM &-1.573 &-1.549 &-0.846  \\
    &-0.899 &-0.863 &-0.807\\
    &-1.154 &-1.093 &-0.758\\
\colrule
CBM &-0.105 &-0.105 &1.372\\
    &-0.106 &-0.105 &1.359\\
\end{tabular}
\end{ruledtabular}
\end{table}

In order to see the contribution from different states to the electronic structure of FeVSb partial density of states (DOS) along with total DOS (TDOS) are calculated. The obtained TDOS in states/eV/f.u. and partial DOS (PDOS) in states/eV/atom are shown in Fig. 1 (c). In the figure $E_{F}$  is set to the middle of the gap. As can be seen from the figure, in FeVSb the electronic states in valence band region ($\sim$ -2 eV to $\sim$ -0.15 eV) are mainly contributed from Fe $3d$ states. While in the conduction band region ($\sim$0.16 eV to $\sim$2 eV) the predominant contribution is seen by V $3d$ states compared to Fe $3d$ states. This indicates  that states which significantly contribute to transport properties are mainly governed by the Fe and V transition elements in FeVSb. Therefore, in tuning the electronic structure to get desired electronic transport properties by doping one can substitute the Fe or V element. The contribution of Sb 5$p$ states to the electronic DOS in the $\sim$ -2 eV to $\sim$3 eV region is very less compared to that from either Fe or V. Thus one may expect less changes in the electronic transport properties in doping Sb position. But, this also hints that one can dope Sb position with other heavy elements to modify the phonon vibrations thereby less affecting the electronic dispersion near the boundaries of band gap. This is useful in reducing the lattice thermal conductivity of the compound. The Seebeck coefficient $S$ and other transport properties of a thermoelectric material depend on the band gap and features of the electronic dispersion in the valence and conduction band region. Using the electronic structure with SOC discussed above we try to give an explanation for the reported experimental $S$ of two FeVSb samples with the help of transport calculations under Boltzmann transport theory.

\subsection{\label{sec:level2}Seebeck coefficient}
The Fig. 2 shows the experimental $S$ of two FeVSb samples along with the calculated ones. The experimental data of $S$ are taken from the work of Fu et al. \cite{fu2012,fu2013} The nature of the experimental $S$ can be observed in Fig. 2 (a) \cite{fu2013} and 2 (b) \cite{fu2012}, which are labeled with symbols S1 and S2, respectively in the plot. The magnitude of experimental $S$ of both samples are increasing upto the near middle of temperature range and start to decrease after reaching peak values of $\sim$ -263 $\mu VK^{-1}$ (S1) and $\sim$ -266 $\mu VK^{-1}$ (S2) at 550 K and 500 K, respectively. In order to understand this behaviour of $S$, initially question arises what might be the band gap of the FeVSb. Knowing the right band gap is important in the calculation of $S$ since it mainly decides the carrier concentration and hence the $S$ value. Also, there is no reported experimental value of band gap from optical measurements in the literature to the best of our knowledge to use in the theoretical calculations. This leads one to consider the theoretical value of band gap in the calculation of $S$ of this compound. But, the theoretical values of band gap calculated under different methods are varying as discussed earlier.  Therefore, question arises what may be the right band gap of FeVSb which can better explain the experimental $S$. Therefore, under rigid band approximation, the band gap values are varied in the range 0.2 eV to 0.7 eV and $S$ values are calculated. In calculating the $S$, the temperature dependence on the chemical potential was included using the below relation \cite{ashcroft,paper4},
\begin{equation}
\mu = E_{v} + \frac{1}{2}E_{g} + \frac{1}{2}k_{B}T ln \left(\frac{{\sum\limits_{\alpha}}(m^{\alpha}_{v})^{3/2}}{{\sum\limits_{\alpha}}(m^{\alpha}_{c})^{3/2}}\right).
\end{equation}
Here, $k_{B}$ is Boltzmann constant, $\mu$ is chemical potential, $E_{c}$ ($E_{v}$) is the energy at the CBM (VBM). The term $m^{\alpha}_{c}$ ($m^{\alpha}_{v}$) is the geometric mean of the eigenvalues of effective mass tensor at CBM (VBM), where $\alpha$ denotes the index of the degenerate band at CBM or VBM. The values of $m^{\alpha}_{c}$ ($m^{\alpha}_{v}$) are calculated from the eigenvalues of effective mass tensors from Table I. The above relation is employed to further calculate the temperature dependent $\mu$ values in FeVSb. 

We calculated the $S$ of FeVSb under Boltzmann transport theory and rigid band approximation. At a given temperature, $S$ as a function of $\mu$ (independent of temperature) is obtained for various band gaps from 0.2 eV to 0.7 eV. Then by applying Eq. 2, temperature dependent chemical potential $\mu(T)$ values are obtained and thus, corresponding $S$ values are calculated. The Seebeck coefficient thus calculated are shown in Fig. 2 (a) and 2 (b), respectively. The best possible matching between calculated and experimental $S$ could be obtained for band gap value of 0.7 eV. The values of $\mu$ at 300 K for samples S1 and S2 at which matching are found to be $\sim$19.7 meV and $\sim$19.3 meV away from the middle gap towards the conduction band region, respectively. The calculated $S$ for both the samples are found to be in qualitatively good agreement with the experimental data as can seen in the figure. Here, it is important to notice that the band gap obtained from mBJ is also $\sim$0.7 eV. The mBJ functional is generally known to give an accurate value of band gap for semiconductors. This indicates that band gap of FeVSb could be $\sim$0.7 eV as this theoretical band gap gives a possibly best explanation of experimental $S$ of the sample .  Also, it is important to note that the experimental $S$ of FeVSb reported in the literature are varying from each other \cite{young,jodin,yamamoto}, with possible presence of two phases in the sample or depending on the sample preparation method. The possible reasons for the observed deviation from the experimental data may be due to temperature dependent band structure and band gap which are not considered here. Also, in the experimental samples normally Sb or other defects may be present which can alter the transport properties compared to that of perfect crystal which is calculated here under rigid band approximation.

\begin{figure}
\includegraphics[width=6cm, height=10cm]{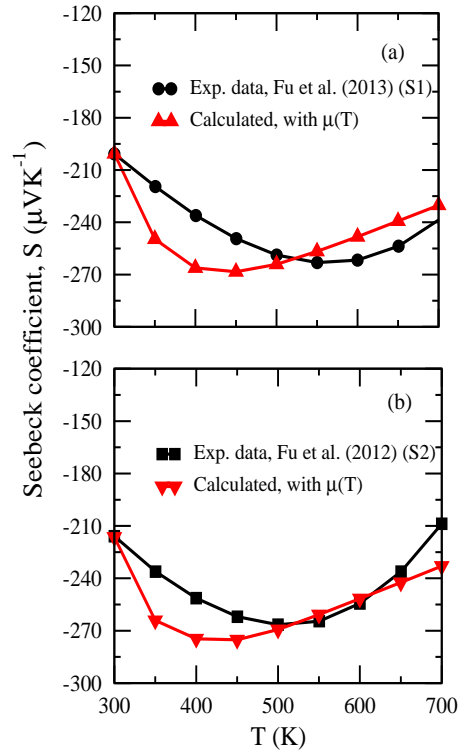} 
\caption{The experimental and calculated Seebeck coefficient for two FeVSb samples. The experimental data S1 in (a) and S2 in (b) are from Ref. \cite{fu2013} and \cite{fu2012}, respectively.  }
\end{figure}

\subsection{\label{sec:level2}Phonon properties}
In this section, we report the phonon dispersion, phonon DOS,  specific heat at constant volume and Helmholtz free energy $F_{ph}$ of FeVSb half-Heusler calculated under harmonic approximation. The calculated phonon dispersion of FeVSb along high symmetric directions is shown in Fig. 3 (a) in which the nonanalytical term correction (NAC) is not included. There are nine branches corresponding to three atoms in the primitive cell. The six optical branches are well separated from the acoustic phonon branches along the directions shown. At the $\Gamma$-point the optical branches become triply degenerate at $\sim$28 meV and $\sim$ 36meV, respectively. The minimum gap between the acoustic and optical branches at the $L$-point is $\sim$3.7 meV. The separation among these branches is getting higher at other $q$-points shown in the figure. 

Further, in order to see the effect of long range Coulomb interaction on the phonon frequencies NAC is included. The Born effective charges (BEC) required for the NAC are calculated. The calculated values of BEC for Fe, V and Sb ions are $\sim$8.1, $\sim$ -5.5, $\sim$ -2.6, respectively. The dielectric constant value needed for NAC is taken from Ref. \cite{nac}. The calculated phonon dispersion by including NAC is shown in Fig. 3 (b). Due to NAC, the lifting in the degeneracy of optical phonon branches compared with Fig. 3 (a) can be observed at the $\Gamma$-point. This suggests the importance of including NAC in the further phonon properties calculations. But, this effect is found to be negligible at other part of the dispersion as can be observed by comparing with Fig. 3 (a). These theoretical observations can be compared and verified if one performs experimental measurement of the phonon dispersion.

Further, phonon total DOS along with partial DOS are calculated to see the contribution of vibrational modes from different atoms to the acoustic and optical phonons. The calculated phonon total DOS per primitive cell and partial DOS per atom  are shown in Fig. 4. As can be seen from the figure there is a gap in the DOS around 25 meV, separating the states corresponding to acoustic and optical phonons. The lower energy acoustic phonon modes (below 25 meV) are predominantly due to the vibrations of heavier mass Sb atom. The optical phonons with energy in the range $\sim$26 meV to $\sim$34 meV are mainly from the vibrations of V atom. While, for the higher energy optical phonons (above $\sim$ 34 meV) mainly vibrations of Fe atom contribution can be observed. 

The maximum frequency of phonon is one of the measures of Debye frequency $\omega_{D}$. From the total DOS in Fig. 4 one can observe that $\omega_{D}$ is $\sim$47 meV in case of FeVSb. One can estimate the Debye temperature $\Theta_{D}$ from the value of $\omega_{D}$ using the relation $\hbar \omega_{D}=k_{B} \Theta_{D}$. The calculated $\Theta_{D}$ in case of FeVSb is $\sim$547 K. The Debye temperature value in case of FeVSb is higher compared to that of the half-Heusler thermoelectric ZrNiSn ($\sim$382 K) \cite{paper4}. Further, the phonon contribution to the constant volume specific heat $c_{v}$ and Helmholtz free energy $F_{ph}$ are calculated under harmonic approximation. The inclusion of NAC is found to be less significant on the calculated thermodynamical properties $c_{v}$ and $F_{ph}$. The calculated $c_{v}$ and $F_{ph}$ as a function of temperature are shown in Fig. 5, along with the experimental constant pressure specific heat $c_{p}$ from the work of Fu \textit{et al.} \cite{fu2012}.  The variation in the value of $c_{v}$ is almost constant above 500 K and approaching the classical Dulong-Petit limit of $\sim$75 J mol$^{-1}$K$^{-1}$ at high temperatures. The calculated $c_{v}$ is qualitatively close to the nature of $c_{p}$ with the change in temperature. The values of $c_{v}$ are close to the experimental $c_{p}$ upto $\sim$200 K. The nature of phonon contribution to $F_{ph}$ as a function of temperature in case of FeVSb can also be seen from Fig. 5. The intercept of the $F_{ph}$ to the y-axis represents the zero-point energy. The value of zero-point energy in case of FeVSb is $\sim$12 kJ/mol. 

\begin{figure}
\includegraphics[width=8.5cm, height=3.5cm]{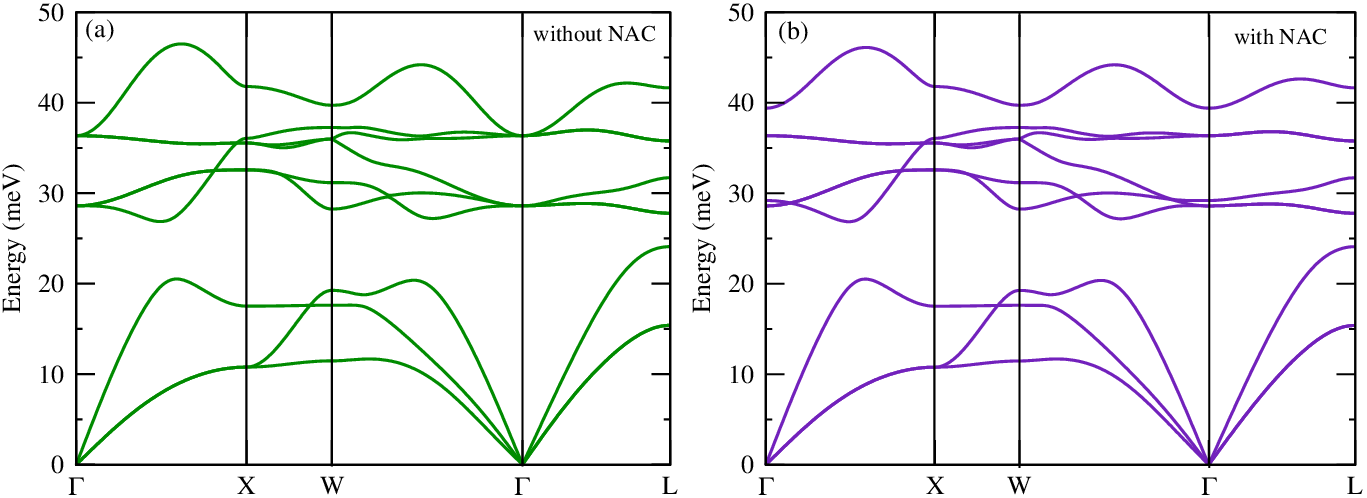} 
\caption{The phonon dispersion of FeVSb. (a) Before including NAC and (b) after including NAC.}
\end{figure}

\begin{figure}
\includegraphics[width=8cm, height=5cm]{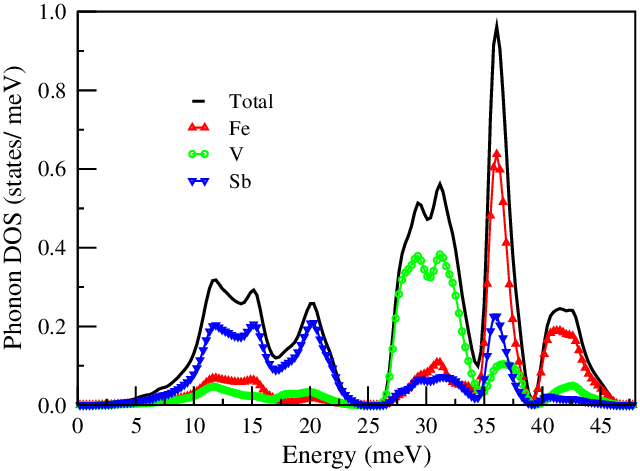} 
\caption{The phonon total DOS per unit cell and partial DOS per atom for FeVSb. }
\end{figure}

\begin{figure}
\includegraphics[width=8cm, height=5cm]{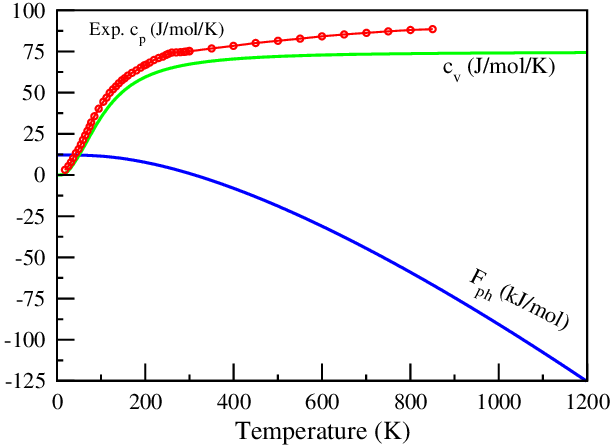} 
\caption{The phonon part of constant volume specific heat $c_{v}$ and Helmholtz free energy $F_{ph}$ along with the experimental $c_{p}$\cite{fu2012} for FeVSb.  }
\end{figure}

\subsection{\label{sec:level2}Thermal expansion}
We report the thermal expansion behaviour in FeVSb compound calculated under quasi-harmonic approximation (QHA) \cite{phonopy}. For a thermoelectric material like FeVSb with promising application potential, knowing the thermal expansion behaviour is desirable. This would be useful in designing the thermoelectric generator (TEG) according to its working temperature range in order to avoid possible thermal stress (or fatigue). Thus, to see the nature of thermal expansion in FeVSb, the linear thermal expansion coefficient is calculated. The calculated linear thermal expansion coefficient $\alpha (T)$ of FeVSb is shown in Fig. 6 (c). The Fig. 6 (a) and (b) show the variation of total free energy $F$ with primitive cell volume at different temperatures and change in the equilibrium volume of primitive cell with temperature, respectively. The total free energy $F$ at a given volume $V$ is obtained as the sum of the relative ground state electronic total energy $U_{el}(V)-U_{el}(V_{0})$ (where $V_{0}$ is equilibrium volume $V_{0}$ at 0 K) and the phonon contribution to Helmohltz free energy $F_{ph}(T;V)$. Therefore, the $F$ in Fig. 6 (a) is calculated as, $F=[U_{el}(V)-U_{el}(V_{0})]+F_{ph}(T;V)$. The minimum free energy point corresponding to equilibrium volume at each temperature is obtained after fitting the Birch-Murnaghan equation of states \cite{birch} (EOS) to the $F$ \textit{vs.} Volume curves. Each of the such equilibrium volume (minimum free energy) point at a given temperature is connected by the solid (red) line in Fig. 6 (a). This line gives the change in the equilibrium volume of the primitive cell with temperature. This variation of volume as a function of temperature upto 1200 K is shown in Fig. 6 (b). The volume at temperature of 1200 K is changed by $\sim$2.5 \% with respect to its ground state volume.

In Fig. 6 (c), the calculated $\alpha (T)$ of FeVSb with change in temperature upto 1200 K is presented. For a cubic crystal assuming uniform expansion in three directions, the $\alpha (T)$ can be obtained from the volumetric thermal expansion coefficient $\beta (T)$ as $\alpha (T)=\frac{1}{3}\beta (T)$ \cite{ashcroft}. Here, the term $\beta (T)$ is defined as $\beta(T)= \frac{1}{V(T)} \frac{\partial V(T)}{\partial T}$. From the figure one can see the sharp increase in the value of $\alpha (T)$ upto temperature of $\sim$260 K which indicates high rate of change in the volume in the crystal. Above 250 K the rate of change is reducing upto $\sim$440 K and after this temperature the $\alpha (T)$ is increasing at almost constant rate. The $\alpha (T)$ is found to vary as $\sim T^{3}$ upto $\sim$20 K and as $\sim$ constant at high temperatures showing nearly the same temperature dependence as $c_{v}$ at the two extreme cases \cite{ashcroft}.  The value of $\alpha (T)$ at 300 K is 6.9x10$^{-6} K^{-1}$ which is increasing with temperature and reaches a value of 8.4x10$^{-6} K^{-1}$ at 1200 K. We could not come across any report of the experimental thermal expansion of FeVSb in the literature. Therefore, it is desirable to verify our theoretical calculations under QHA with the experimental measurements.

\begin{figure*}
\includegraphics[width=14cm, height=7cm]{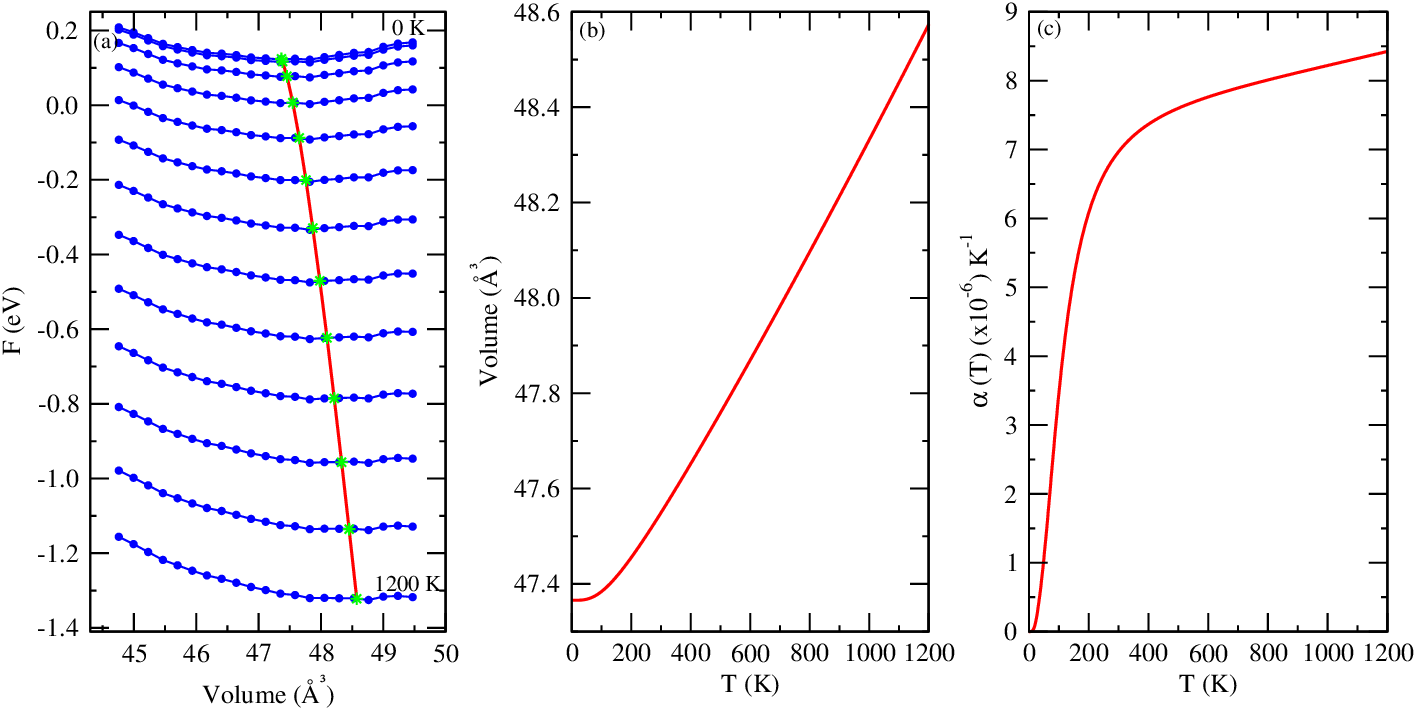} 
\caption{(a) Variation of total free energy $F$ with primitive cell volume. (b) Variation of primitive cell volume with temperature T. (c) Linear thermal expansion coefficient $\alpha(T)$ with temperature T for FeVSb.}
\end{figure*}

\subsection{\label{sec:level2}Lattice thermal conductivity}
In this section, we report and discuss the temperature dependent lattice part of thermal conductivity  $\kappa_{ph}$ of FeVSb calculated using anharmonic first-principles phonon calculations. The lattice thermal conductivity under single-mode relaxation time (SMRT) approximation by solving linearized Boltzmann transport equation (LBTE) method is given by \cite{phono3py},
\begin{equation}
\kappa_{ph} = \frac{1}{NV_{0}} {\sum\limits_{\lambda}} c_{\lambda} {\bf{v}_\lambda} \otimes {\bf{v}_\lambda} \tau_{\lambda}^{SMRT}.
\end{equation}
Here, $N$ is number of unit cells in the crystal, $V_{0}$ is volume of the unit cell, $c_{\lambda}$ is mode dependent specific heat, ${\bf{v}_\lambda}$ is the group velocity of phonon mode and $\tau_{\lambda}^{SMRT}$ is the single-mode relaxation time of phonon mode $\lambda$. Here, the symbol $\lambda$ denotes a phonon mode with wave vector $\textbf{q}$ and branch index $j$. The phonon relaxation time, $\tau_{\lambda}^{SMRT}$ is approximately taken to be the life of phonon mode, $\tau_{\lambda}$. The phonon lifetime of mode $\lambda$ is given by \cite{phono3py},
\begin{equation}
\tau_{\lambda} = \frac{1}{2\Gamma_{\lambda}(\omega_\lambda)},
\end{equation}
where, $2\Gamma_{\lambda}(\omega_\lambda)$ corresponds to the phonon linewidth which is computed using the anharmonic third order force constant \cite{phono3py}. 
 
\begin{figure*}
\includegraphics[width=15cm, height=6cm]{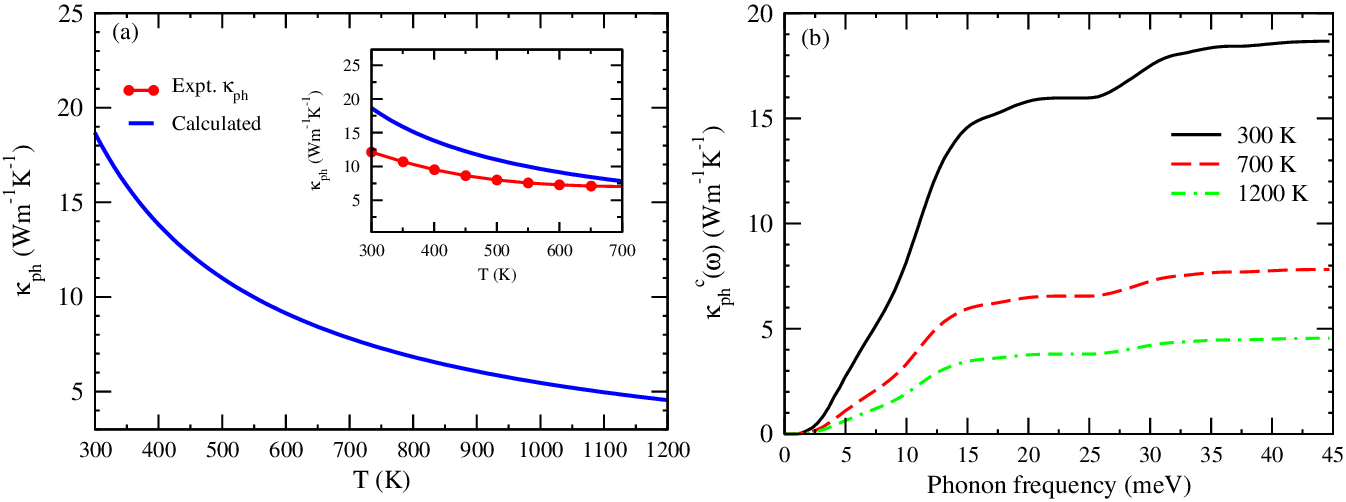} 
\caption{(a) The calculated lattice part of thermal conductivity $\kappa_{ph}$ of FeVSb with temperature. The inset shows the comparison of experimental $\kappa_{ph}$ \cite{fu2012} with the calculated $\kappa_{ph}$. (b) The calculated cumulative lattice thermal conductivity $\kappa_{ph}^{c}(\omega)$ as a function of frequency (energy) at three temperatures.}
\end{figure*}

The calculated $\kappa_{ph}$, in cubic FeVSb as a function of temperature in the range 300-1200 K is presented in Fig. 7 (a). The nonanalytical term correction was included in the calculation of $\kappa_{ph}$ as we have observed earlier that, it changes energy of the phonons around $\Gamma$-point. The inset of Fig. 7 (a) shows the comparison between the calculated and extracted experimental lattice thermal conductivity (denoted as Expt. $\kappa_{ph}$ in figure) as given in the work of Fu \textit{et al.} \cite{fu2012}. The calculated $\kappa_{ph}$ has decreasing nature in the temperature range shown. The calculated value of $\kappa_{ph}$ at 300 K is $\sim$18.6 W$m^{-1}K^{-1}$ and it reaches value of $\sim$7.8 W$m^{-1}K^{-1}$ and $\sim$4.5 W$m^{-1}K^{-1}$ at 700 K and 1200 K, respectively. While, the respective experimental value of $\kappa_{ph}$ at 300 K and 700 K are $\sim$12.1 W$m^{-1}K^{-1}$ and $\sim$7 W$m^{-1}K^{-1}$. Carrete \textit{et al.} have estimated the room temperature lattice thermal conductivity of set of half-Heuslers including FeVSb \cite{jesus}. The value of lattice thermal conductivity in their work is 24.1 W$m^{-1}K^{-1}$ (at 300 K) which is higher than the obtained value in our work. The calculated values of $\kappa_{ph}$ above 500 K are close to the experimental value. But, below 500 K calculated values are quite higher compared to the experimental values of $\kappa_{ph}$. In the figure, a reduction in the separation between the calculated and the experimental $\kappa_{ph}$ curves can be observed as temperature approaches higher values. The deviation of the calculated $\kappa_{ph}$ compared to the experimental value of $\kappa_{ph}$ can be normally attributed to many reasons. For instance, generally in the polycrystalline sample defects and/or disorder will be present. Also, the grain size and sample preparation condition are two of the many important factors which affect the thermal conductivity of sample. Here in the calculation of the $\kappa_{ph}$ of single crystalline FeVSb, only phonon-phonon interactions are considered while in the real solid, the relaxation time also depends on the phonon-electron and phonon-defect interactions. This can be the reason for the higher calculated values of $\kappa_{ph}$ of FeVSb which may further reduce by including phonon-electron interactions. Therefore, it is desirable to compare the calculated values with the single crystalline experimental data. 

Further, to see the amount of contribution of the phonon modes to total lattice thermal conductivity at a given temperature, the cumulative lattice thermal conductivity $\kappa_{ph}^{c}(\omega)$ as a function of frequency (energy) is calculated. The cumulative lattice thermal conductivity is defined as \cite{kappacum},
\begin{equation}
\kappa_{ph}^{c}(\omega)= \int_{0}^{\omega} \kappa_{ph} \delta (\omega^{\prime}-\omega_{\lambda}) d\omega^{\prime},
\end{equation}
where, the term $\kappa_{ph}$ is given by the Eq. 3. The calculated $\kappa_{ph}^{c}(\omega)$ for FeVSb at three temperatures 300K, 700K and 1200 K are shown in Fig. 7 (b). At a given temperature, the value of $\kappa_{ph}^{c}(\omega)$ at the highest energy of $\sim$45 meV gives the total $\kappa_{ph}$ at that temperature. The values of $\kappa_{ph}$ at 300 K, 700 K and 1200 K are $\sim$18.6, $\sim$7.8, and $\sim$4.5 W$m^{-1}K^{-1}$, respectively.   These values can be observed by looking at the value of $\kappa_{ph}$ and $\kappa_{ph}^{c}(\omega)$ from Fig. 2 (a) and (b), respectively for the temperatures mentioned. As we have discussed in subsection III. C, the acoustic and optical phonons are separated by a clear gap around $\sim$25 meV. This can be observed from the total phonon DOS shown in Fig. 4. The modes with energy upto $\sim$24 meV comprise of acoustic phonons while the optical modes can be observed above energy of $\sim$26 meV. Considering the distribution of acoustic and optical phonons in the energy range, we estimate an approximate percentage contribution to $\kappa_{ph}$ from phonon modes using Fig. 7 (b). The estimated contribution to $\kappa_{ph}$ from acoustic phonons are $\sim$85.6, $\sim$83.7 and $\sim$83.4 \% at 300 K, 400 K and 1200 K, respectively. This indicates that acoustic phonons are playing dominant role in the lattice thermal conductivity behaviour in FeVSb half-Heusler. While the contribution from optical phonons are less compared to acoustic phonons. This may be due to the shorter lifetime of optical phonons. So, it is interesting to further study and analyse the lifetimes of acoustic and optical phonons. The calculated $\kappa_{ph}$ can be useful in the calculation of figure of merit and in predicting the temperature dependent thermoelectric properties.

\subsection{\label{sec:level2}Figure of merit and efficiency}
\begin{figure*}
\includegraphics[width=10cm, height=8cm]{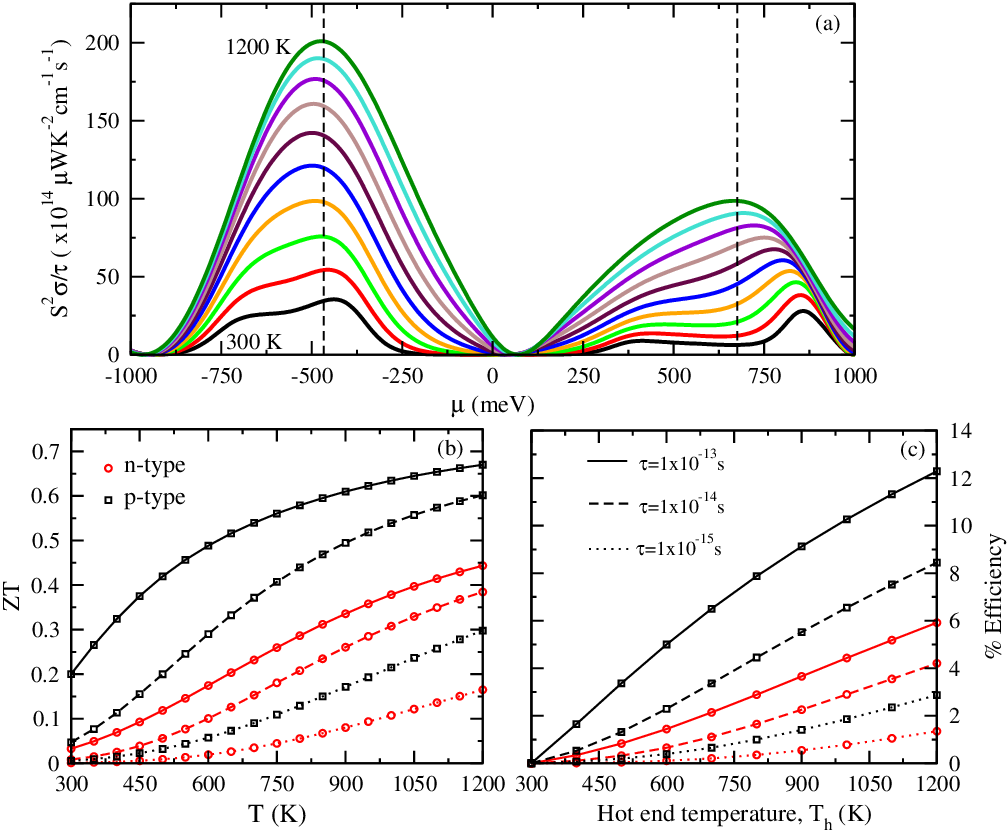} 
\caption{(a) Power factor per relaxation time ($S^{2}\sigma /\tau$) with change in chemical potential $\mu$. (b) Figure of merit $ZT$ and (c) \% Efficiency ($\eta$) as a function of temperature. }
\end{figure*}
The measure of thermoelectric performance of a material is given by its figure of merit $ZT$ and efficiency $\eta$. In this section, a prediction of $ZT$ and \% efficiency possibly achievable in FeVSb half-Heusler are given. One of the routes of enhancing the $ZT$ is by improving the power factor ($S^{2}\sigma$) obtainable from that material. The power factor of a material is dependent on the electronic band structure. Thus, the power factor can be enhanced by figuring out the suitable carrier concentration from that electronic structure. In order to find out the optimal carrier concentration, we calculated power factor per relaxation time $S^{2}\sigma /\tau$ (PF) as a function of chemical potential $\mu$. In the Boltzmann transport calculations, the doping of carriers is taken care through the change in the $\mu$. Thus, by calculating PF as a function of $\mu$ one can get an estimate of the carrier concentration corresponding to maximum power factor. 

The calculated PF in $10^{14}\mu WK^{-2}cm^{-1}s^{-1}$ (PFU) is shown in Fig. 8 (a). In the figure, the negative values of $\mu$ correspond to the p-type doping while the positive values correspond to n-type doping. As can be seen from the figure, there are two peaks of PF at a given temperature. These peaks correspond to the maximum PF obtainable at that temperature. Also, it can be observed that PF due to p-type doping is higher compared to that of n-type doping. The values of $\mu$ at which maximum PF observed for n-type and p-type doping are $\sim$675 meV and  $\sim$ -466 meV, respectively. These peaks are marked with dashed lines in the figure. The consecutive temperature has to be read in the order of the peak height. The value of maximum PF at 1200 K are $\sim$98 and $\sim$200 PFU for n-type and p-type doping, respectively. This value of maximum PF due to hole doping obtainable in FeVSb is higher compared to that in ZrNiSn studied in our previous work \cite{paper4}. Suggesting the value of optimal carrier concentrations for doping is useful rather than the $\mu$ values to realize the reported thermoelectric properties. Therefore, the electron and doping concentration for these $\mu$ levels are calculated. The value of optimal hole concentration (p-type doping) calculated that yields maximum PF in FeVSb is $\sim$1.5x10$^{21}cm^{-3}$. Similarly, the value of optimal electron concentration (n-type) is found to be $\sim$1.7x10$^{21}cm^{-3}$.

Further, to evaluate the material for thermoelectric application, the $ZT$ and efficiency are calculated for the doped FeVSb with optimal electron and hole concentrations. In the 	calculation of $ZT$, relaxation time ($\tau$) values of 1x10$^{-13}$s ($\tau_1$), 1x10$^{-14}$s ($\tau_2$) and 1x10$^{-15}$s ($\tau_3$) which normally correspond to semiconducting, metallic and defective-sample regimes, respectively are considered. For the calculation of $ZT$, the $\kappa_{ph}$ are taken from section III. E and used. The obtained $ZT$ in the temperature range 300-1200 K can be seen in Fig. 8 (b). The different $\tau$ values using which the $ZT$ is calculated are indicated by the line style mentioned in Fig. 8 (c). The values of $ZT$ for p-type doping is higher compared to n-type doping for a particular $\tau$.  The $ZT$ has increasing nature with the increase in temperature. 
For p-type doping the maximum $ZT$ at 1200 K are $\sim$0.66, $\sim$0.60 and $\sim$0.28 for $\tau_1$, $\tau_2$ and $\tau_2$, respectively. Similarly, the n-type doping gives the maximum $ZT$ of $\sim$0.44, $\sim$0.38 and $\sim$0.16 at 1200 K for the three consecutive $\tau$ values used respectively.

The obtained $ZT$ for doped FeVSb are fairly high upto used $\tau$ value of 1x10$^{-14}$s. But, here it is important to note that in the calculation of $ZT$, $\kappa_{ph}$ of the pure FeVSb is used. This $\kappa_{ph}$ value will be higher compared to that of the doped FeVSb. As reported in the literature, one can systematically reduce $\kappa_{ph}$ by doping with heavy elements \cite{fu2012,shenexp}. Thus, if we consider lower values of $\kappa_{ph}$ the $ZT$ can be further enhanced. Normally, in the high temperature region, for half-Heuslers, after doping with heavy elements the $\kappa_{ph}$ can be reduced by $\sim$25\% of its parent compound depending upon the doping amount and element. So, considering this reduction in calculated $\kappa_{ph}$ of FeVSb, the possible enhancement in $ZT$ is estimated. This led to a $ZT$ of $\sim$0.35 (p-type) and $\sim$0.19 (n-type) at 1200 K for $\tau$ of 1x10$^{-15}$s. Further, we tried Co, Ni, Cu, Zn, Sc, Ti elements for doping and Sb vacancy for the possibility of achieving n-type or p-type FeVSb through KKR-CPA \cite{akaikkr} calculations. Out of these possibilities tried, Sb vacancy (1 \%) is found to give p-type levels while, Zn doping (3 \%) is found to give n-type levels in FeVSb. This is shown for Sb vacancy and Zn doping in representative TDOS plots in Fig. 9 (a) and (b), respectively. These possibilities can be tried in appropriate concentration to achieve optimal power factor.

Once, the $ZT$ and working temperature of a thermoelectric material are known, it is desirable to calculate its conversion efficiency. This would give an estimate for the applicability in single-stage or hybrid thermoelectric generator (TEG). Here, we applied the segmentation method given by Gaurav \textit{et al.} \cite{gaurav} to calculate the efficiency of p-type and n-type thermoelectric materials with optimal carrier concentration. In calculating the efficiency, the temperature difference ($\Delta T$) of 10 K across each segment is considered. The cold end temperature $T_c$, is fixed at 300 K and hot end temperature $T_h$ is varied in steps of 100 K upto 1200 K. The calculated \% efficiency for p-type (square symbol) and n-type (circle symbol) FeVSb for three relaxation time values mentioned are shown in Fig. 8 (c). The \% efficiency for p-type FeVSb is higher compared to n-type FeVSb for a given value of $\tau$. The increase in the \% efficiency with increase in the $T_h$ can be observed in the figure. For the p-type FeVSb, the \% efficiency when $T_h$ is 1200 K are, $\sim$12.2, $\sim$8.4 and $\sim$2.7 \%, for three consecutive $\tau$ values mentioned, respectively. Similarly, the n-type FeVSb shows \% efficiency of $\sim$6.0 ($\tau_1$), $\sim$4.2 ($\tau_2$) and $\sim$1.3 ($\tau_3$) \%, for $T_h$ of 1200 K. These efficiency can be further enhanced if $\kappa_{ph}$ of doped FeVSb is used in the calculation.
\begin{figure}
\includegraphics[width=8.5cm, height=3.5cm]{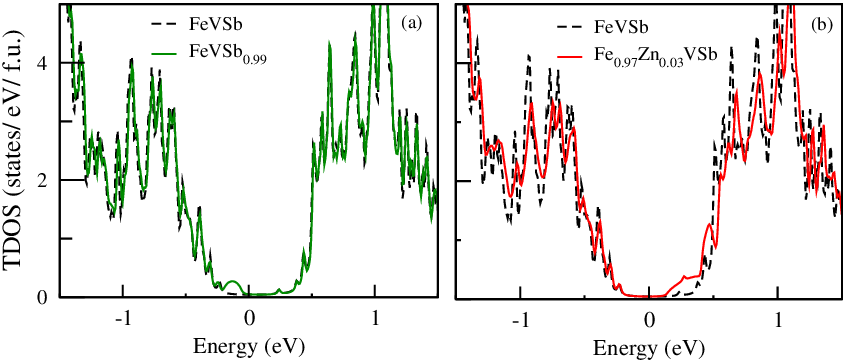} 
\caption{The TDOS of (a) FeVSb$_{0.99}$ (b) Fe$_{0.97}$Zn$_{0.03}$VSb showing p-type and n-type levels along with the pure FeVSb.}
\end{figure}

\section{Conclusions} 
In this work, the electronic structure of FeVSb is studied using FP-LAPW method with SCAN meta-GGA. The mBJ functional employed is found to give a band gap of $0.7$ eV for FeVSb. The qualitative agreement between the experimental and calculated $S$, suggested that the band gap of FeVSb samples   \cite{fu2012,fu2013} could be $\sim$0.7 eV which is also supported by mBJ calculation. The phonon dispersion, DOS of FeVSb are reported for FeVSb. The effect of long range Coulomb interactions is considered by NAC which is found to lift the degeneracy of phonon modes around $\Gamma$-point. The acoustic and optical phonons are found to be separated by a gap of $\sim$3.7 meV. The phonon PDOS suggested that the acoustic phonon modes are  mainly due to the vibrations of heavy Sb atom. The thermal expansion behaviour in FeVSb is reported under QHA. The volume of crystal is found to change by $\sim$2.5 \% at 1200 K with respect to ground state volume. The $\kappa_{ph}$ is calculated using first-principles anharmonic phonon calculations considering phonon-phonon interaction under single mode relaxation time  approximation in 300 - 1200 K . Qualitatively, the nature of $\kappa_{ph}$ agrees with the experimental $\kappa_{ph}$ in the 300 to 700 K. The obtained $\kappa_{ph}$ at 300 K of $\sim$18.6 W$m^{-1}K^{-1}$ is higher compared to the experimental value of $\sim$12.1 W$m^{-1}K^{-1}$. Above 500 K calculated $\kappa_{ph}$ is close to the experimental values. The cumulative $\kappa_{ph}$ analysis showed that acoustic phonons contribute mainly to the $\kappa_{ph}$ in this compound. The p-type FeVSb is predicted to have higher power factor, $ZT$ and efficiency compared to that of n-type FeVSb. At 1200 K, the maximum predicted $ZT$ is $\sim$0.66 and $\sim$0.44 for p-type and n-type FeVSb, respectively. Similarly, a maximum efficiency of $\sim$12.2 \% and $\sim$6.0 \% is expected for the hot and cold temperature of 1200 K and 300 K, respectively.

\section{Acknowledgements}
The authors thank Science and Engineering Research Board (SERB), Department of Science and Technology, Government of India for funding this work. This work is funded under the SERB project sanction order No. EMR/2016/001511.

\section{References}
\bibliography{ref}
\bibliographystyle{apsrev4-1}

\end{document}